\documentclass[prb,twocolumn,amsmath,amssymb]{revtex4}

\usepackage{graphicx}
\usepackage{dcolumn}
\usepackage{bm}

\begin{document}
\title{Ultrafast dynamics of surface electromagnetic waves in nanohole array on metallic film
}
\author{A. S. Kirakosyan,$^{1,2}$ M. Tong,$^3$ T. V. Shahbazyan,$^1$ and Z. V. Vardeny$^3$} 

\affiliation{$^1$Department of Physics, Jackson State University, Jackson, MS
39217 USA\\
$^2$Department of Physics, Yerevan State University, Yerevan, 375025 Armenia\\
$^3$Department of Physics, University of Utah, Salt Lake City, Utah 84112 USA
}
\begin{abstract}
We study the ultrafast dynamics of surface electromagnetic waves photogenerated on aluminum film perforated with subwavelength holes array by means of transient photomodulation with $\sim 100$ fs time resolution. We observed a pronounced \textit{blueshift} of the resonant transmission band that reveals the important role of plasma attenuation in the dynamics and that is inconsistent with plasmon-polariton mechanism of extraordinary transmission. The transient photomodulation spectra were successfully modeled within the Boltzmann equation approach for the electron-phonon relaxation dynamics, involving non-equilibrium hot electrons and quasi-equilibrium phonons.
\end{abstract}

\maketitle

\section{Introduction}
On a smooth metal-dielectric interface light does not couple to elementary electronic excitations of the metal surfaces because of energy and momentum conservation. But in a metal film that is perforated with a two-dimensional (2D) periodic array of holes forming a 2D metallo-dielectric photonic crystal, the periodicity promotes zone folding that results in the formation of band structure. The grating makes it possible for light to couple directly to surface electromagnetic waves by transfering them the reciprocal lattice momentum $g=2\pi/a$,  where $a$ is the grating period \cite{raether-88}.  Indeed, it was recently found\cite{ebbesen98,schroter98,ghaemi98}  that the optical transmission through hole arrays fabricated on optically thick metallic films is enhanced by orders of magnitude at resonances; this phenomenon has been dubbed "extraordinary transmission" (EOT). The maxima in the zero-order transmission spectrum is alternated by deep minima caused by Wood's anomalies formed due to the diffraction grating of the corrugated film. The optical properties of various subwavelength hole array films have been widely studied and different models have been proposed to explain the underlying mechanism of the anomalous transmission. While earlier models atributed the enhancement to excitation of film's surface plasmon-polaritons (SPP) \cite{ebbesen98,schroter98,ghaemi98,ebbesen01,barnes03}, more recent works adrresed the role resonant cavity modes \cite{porto99,garcia03,lalanne00}, dynamical diffraction resonances \cite{treacy99,treacy02}, and interference of diffracted evanescent waves\cite{lezec04}.  While the precise mechanism of extraordinary transmission is still under discussion, the transient dynamics of surface electromagnetic modes might shed more light on the origins of this phenomenon. 

Here we investigate the ultrafast dynamics of surface electromagnetic modes measured by using  the pump-probe correlation spectroscopy of the transmission bands in 2D subwavelength hole array on aluminum (Al) thin film. We found significant difference with the dynamics observed previously in smooth metal films (mostly gold and silver) in both the time-resolved and spectrally-resolved signals. Following an instantaneous rise at the onset of the impinging pulse, the transient differential transmission exhibits a fast rise on subpicosecond timescale followed by a plateau for time delays of 1-3 ps, and subsequent slow decay with characteristic time of $\sim 40$ ps. The short-time dynamics is accompanied by a blueshift of the main transmission band in the transient spectra that persists for time delays as long as 100 ps. The observed dynamics can be explained by a fast energy transfer in the Al film from the electron gas to the lattice, with subsequent cooling of the Al film by heat transfer to the glass substrate. The fast lattice temperature rise is caused by strong electron-phonon interaction in Al, which makes the electron-lattice energy transfer rate comparable to the rate of nonequilibrium electrons thermalization via electron-electron interactions\cite{tong-prl08}. Importantly, the blueshift is incosistent with the SPP mechanism and points to the interference as the source of the extraordinary transmission.

\section{Experiment}
The plasmonic lattice sample was a 70 nm thick Al film, 5 $\times$ 5 mm$^2$ in area deposited on a glass substrate. The metal film was perforated with circular holes having diameter $D = 150$ nm in a square hole array with lattice constant  $a = 300$ nm resulting in a fractional aperture area of 25\%. The ultrafast laser system used for measuring the transient PM spectrum was a Ti:sapphire regenerative amplifier with pulses of 100 fs duration at photon energies of 1.55 eV, with 400 $\mu$J energy per pulse at a repetition rate of 1 kHz. The second harmonic of the fundamental pulses at 390 nm (3.1 eV) were used as the pump beam. The probe beam was a white light super-continuum generated in a 1-mm thick sapphire plate that covers the spectral range from 1.6 to 2.8 eV; the white light super-continuum was also used for measuring the unperturbed transmission spectrum  $T_0(\omega)$. The pump and probe beams were directed onto the perforated Al film from the air side to maximize coupling of light to the SEW. The transient PM spectrum was obtained from the photoinduced change ($\Delta T$) in $T_0$ using a phase-sensitive technique with a resolution $\Delta T/T_0\sim 10^{-4}$  that corresponds to a photoexcited electron density in the Al metal of $\sim 10^{17}$ cm$^{-3}/$pulse.  
\begin{figure}
\begin{center} 
\includegraphics[width=0.8\columnwidth]{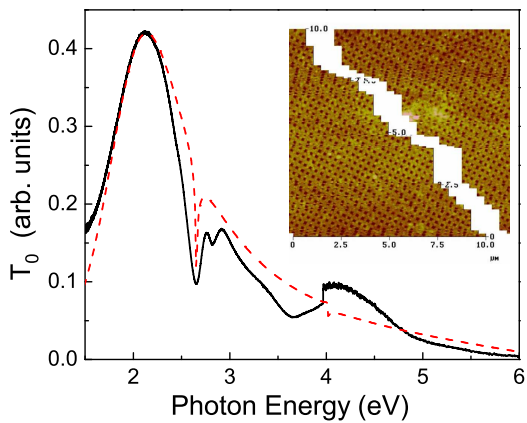}
\caption{The measured (solid line) and calculated (dashed line) normal incidence optical transmission, $T_0$ through an Al plasmonic lattice comprised of a square array of nanoholes with lattice constant 300 nm and hole diameter 150 nm. The inset shows the sample AFM image. Numerical fit is based on the effective medium model for linear transmission  in Sec. \ref{sec:effective}.
}
\label{fig:1}
\end{center}
\end{figure}

Figure \ref{fig:1} shows the EOT spectrum of the Al plasmonic lattice at normal incidence. Several EOT bands and transmission anti-resonances (AR) are observed.  We note that  $T_0(\omega)$ at resonance frequencies substantially surpasses the fractional aperture area of 25\%. The lowest frequency EOT band peaks at $\sim 2.1$ eV; whereas the lowest frequency AR feature that is associated with Wood's anomaly at the Al/glass interface with wavevector  $k^{(01)} = g \varepsilon_{gl}^{-1/2}$ (where $\varepsilon_{gl}=1.56$  and $g=2\pi/a$ ) is in the form of a dip at 2.6 eV.  The AR's at higher frequencies are higher-order Wood's anomalies that correspond to wavevectors $k^{(mn)}= g\left( m^2 + n^2\right)^{1/2}\varepsilon_d^{-1/2}$ (\textit{m, n} are integers) as follows:  3.7 eV ($k^{(11)}  = g \varepsilon_{gl}^{-1/2} 2^{1/2}$) that corresponds to the (1,1) branch at the Al/glass interface; and 4.0 eV ($k^{(01)} = g$) that corresponds to the (0,1) branch at the Al/air interface.  It is noteworthy that the lowest frequency EOT resonance (at $\sim 2.1$ eV) occurs at considerably lower (by $\sim 0.6$ eV) frequency than that predicted by the lowest-order SPP with wavevector satisfying the Bragg condition \cite{ebbesen98}
\begin{eqnarray}
\label{bragg}
k_{spp}^{(mn)}  = g \left[ \left( m^2 + n^2\right) \left( \varepsilon_{m}'^{-1} + \varepsilon_{d}^{-1} \right) \right]^{1/2}.
\end{eqnarray}
\begin{figure}
\centering
\includegraphics[width=2.5in]{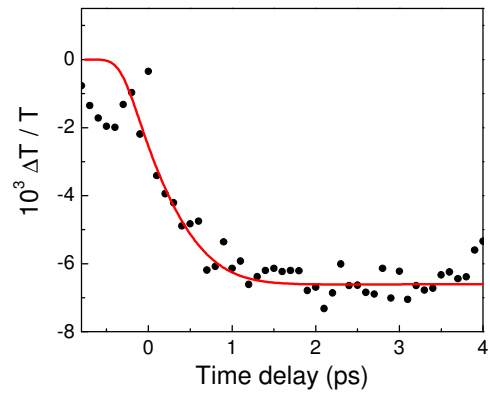}
\caption{ 
Short-time evolution of differential transmission at 1.9 eV is shown together with numerical fit (solid line) based on solution of the Boltzman equation in Sec. \ref{sec:ultrafast}.
} 
\label{fig:2}
\end{figure}

\begin{figure}
\vspace{5mm}
\centering
\includegraphics[width=2.5in]{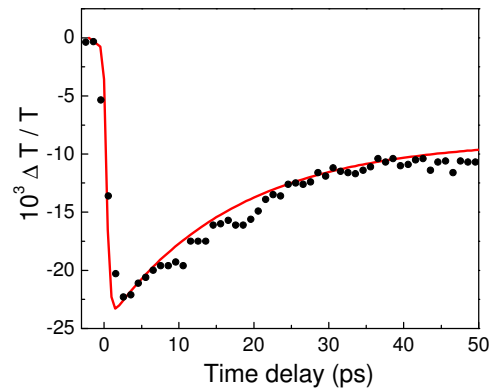}
\caption{ 
Long-time evolution of differential transmission at 1.9 eV due to heat transfer to the glass matrix is shown together with numerical fit (solid line) based on two-temperature model  in Sec. \ref{sec:ultrafast}.
} 
\label{fig:3}
\end{figure}

\begin{figure}
\centering
\includegraphics[width=2.5in]{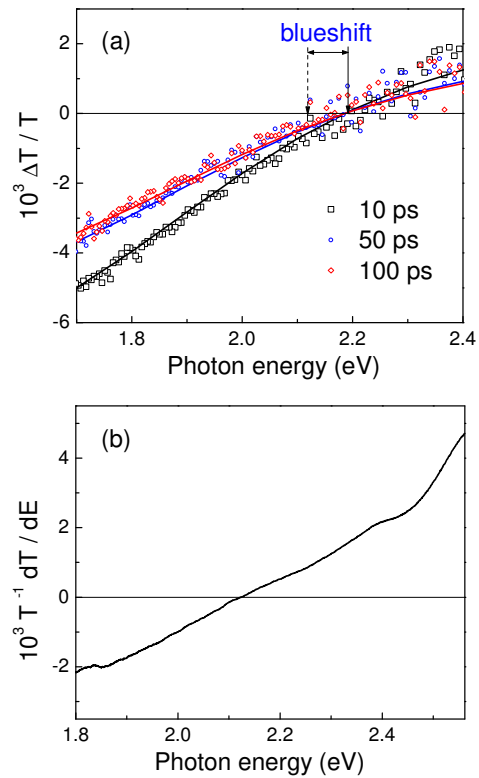}
\caption{ 
(a) Transient spectra at different time delays are shown together with numerical fit (solid lines) based on solution of the Boltzman equation in Sec. \ref{sec:ultrafast}. (b) Log derivative of measured zero-order transmission.
} 
\label{fig:4}
\end{figure}

The transient PM spectra of the perforated Al film are shown in Figs.\ \ref{fig:2} - \ref{fig:4}. The time-evolution of pump-probe signal at 650 nm, corresponding to probe photon energy of 1.9 eV, reveals two component  of the dynamics. Firstly, there is a fast rise of pump-probe signal on timescale $<1$  ps, followed by a plateau at 1-3 ps (see Fig.\ \ref{fig:2}).  Secondly,  after about 3 ps the saturated PM decays with characteristic time of $\sim 40$ ps (see Fig.\ \ref{fig:3}) due the heat transfer to the glass substrate. Figure\  \ref{fig:4}(a) shows the transient PM spectral change from 1.7 to 2.4 eV corresponding roughly to the main transmission band spectral width.  A transient blueshift  is apparent that persists for times up to 100 ps, i e. much longer that electron gas thermalization time, pointing to a prominent role of {\em e-ph} interactions in the dynamics. However this shift cannot be well reproduced by a simple spectral derivative [see Fig.\ \ref{fig:4}(b)] indicating that a perturbation approach cannot properly describe the observed changes in optical spectra for our excitation intensities. At the same time, although the data becomes more noisy at higher energies as one approaches to the Wood anomaly, a clear nonlinearity in the frequency dependence emerges. The energy dependence of $dT/T$ is determined by transient change in the mode dispersion that reflects the periodicity of the structure. 

The short-time dynamics exhibts significant differences with that observed in gold and silver films \cite{sun94,groeneveld95}. In these studies, the picosecond dynamics consisted of two well-separated in time stages: thermalization of nonequilibrium electrons due to electron-electron ({\em e-e}) scatering at 200-500 fs accompanied by the signal rise, followed by the decay on a 1-2 ps  timescale due to the cooling of hot electron gas to the metal lattice. The dominant factor that determines the transient dynamics in noble metals is the change in the interband succeptibilily of the highly polarizable {\em d}-band electrons that is very sensitive to the smearing of the Fermi edge due to the rise in electron temperature $T_e$. In contrast, in simple metals such as Al, the transient optical spectra are governed by the changes in the intraband dielectric function $\varepsilon_m(\omega)=1-\omega_p^2/\omega(\omega+i\gamma)$, where $\omega_p$ is the bulk plasmon energy and $\gamma$ is the optical relaxation rate \cite{palik}.  The absorbance of the film and hence the transient dynamics is related to the deviation of $\gamma$ from its intrinsic value $\gamma_0$ that mainly originates from the change in the  electron-phonon ({\em e-ph}) scattering, while the contributions from {\em e-e}, and impurity scatering are relatively weak \cite{smith82}. Note that the {\em e-ph} interaction constant $G$ in Al is an order of magnitude larger than it is in noble metals (for comparison, $G_{Al} = 3.1 \times 10^{17}$ W/Km$^3$ vs. $G_{Au} = 2 \times 10^{16}$ W/Km$^3$), so that the energy exchange between the electron and phonon subsystems takes place on sub-picosecond times \cite{rethfeld02}. The rapid increase in the lattice temperature $T_L$ and hence in the {\em e-ph} relaxation rate accounts for the fast rise of the pump-probe signal until the equilibrium between electron and phonon subsystems is reached at $\sim 1$ ps. This fast dynamics is superimposed with a slow decay due to the heat transfer from metal to glass substrate resulting in a plateau at 1-3 ps (see Figs.\ \ref{fig:2}-\ref{fig:3}). Importantly, the merging of electron thermalization and electron to lattice cooling stages implies that the short time dynamics is intrinsically nonequilibrium; it is therefore described below within the Boltzmann equation approach.

The observed PM blueshift also indicates that the EOT is sensitive to the imaginary part of permittivity, $\varepsilon_{m}''(\omega)$, for the following reason. Within the Drude model for Al, we have  $\varepsilon_{m}'(\omega) = 1 - \omega_{p}^2/(\omega^2 + \gamma^2)$. With such $\varepsilon_{m}'$, the SPP Bragg condition  Eq.~(\ref{bragg}) leads to the following expression for, e.g.,  (0,1) mode frequency (at $\epsilon_{d}=1$): $\omega_{spp}^{01}\approx \omega_{g}\left [ 1-(\omega_{g}^{2}+\gamma^{2})/2\omega_{p}^{2}\right ]$, where $\omega_{g}=cg$ is the frequency associated with the reciprocal lattice period. Then an increase in $\gamma$  upon pulsed excitation leads to  a weak redshift of the resonance frequency, in disagreement with the data. We thus conclude that the observed spectral blueshift cannot be explained by the present version of the SPP mechanism for the EOT phenomenon \cite{ebbesen98}. At the the same time, an increase in $\gamma$  causes an increase in the \textit{attenuation},  $\varepsilon_{m}''$, which for $\gamma\ll\omega$ can be approximated by  $\varepsilon_{m}''  \approx \gamma \omega_{p}^{2}/\omega^3$. If   $\varepsilon_{m}''$ participates in determining the position and lineshape of the EOT bands, then the increase in  $\gamma$ implies \textit{higher} resonant frequencies (blueshift) in agreement with the data. The linear transmission is modeled below using effective medium approach.

\section{Theory}
The thansient optical spectra in metal films can be described in terms of time-dependent dielectric function $\varepsilon_m(\omega,t)=\varepsilon_m(\omega)+\delta\varepsilon_m(\omega,t)$, where $\varepsilon_m(\omega,t)$ is PM contribution due to a small (compared to ground state) fraction of photoexcited carriers. The differential transmission $\Delta T(t) =T(t)-T_0 $ can be then caculated from the linear transmission $T_0(\omega)$ with $\varepsilon_m(\omega,t)$ \cite{sun94,groeneveld95}. While the time-resolved signal is primarily governed by the electron dynamics in metal film encoded in $\delta \varepsilon_m(\omega,t)$, the transient {\em spectra} are determined by the effect of the latter on the {\em dispersion} of surface electromagnetic modes. Modeling of the dynamics thus involves two distinct tasks: (i) description of the linear transmision through nanohole array for {\em real} metal with absorptive dielectric function, and (ii) simulation of non-equilibrium electron-phonon dynamics in Al film. 


\subsection{Effective medium model for linear transmission}
\label{sec:effective}

We begin with {\em analytical} model for the main transmission band that incorporates the effect of the absorptive part of metal dielectric function. We are interested in the transmission of {\em p}-polarized light normally incident on a metal film perforated with square array of cylindrical nanoholes. We assume that the scatering in the direction normal to the plane of incidence is weak and adopt one-dimentional grating approximation for transverse magenetic and electric fields, $H_y(x,z)$ and $E_x(x,z)$. Magnetic fields  in air (media 1), perforated metal (media 2), and glass (media 3), can be written as
\begin{eqnarray}
\label{H1}
H_{1 y} &=& e^{i k_1 z} + \sum_{n} r_{1 n} \, e^{-i k_{1 n} z + i ng x}, \\
\label{H3}
H_{3 y} &=& \sum_{n} t_{3 n} \, e^{i k_{3 n} (z-h) + i ng x }, \\
\label{H2}
H_{2 y}&=& \sum_{i} H_{i}(x) \Bigl[ 
t_{2 i} \, e^{- \kappa_{i} z} + r_{2 i} \, e^{- \kappa_{i} (h -z)}
\Bigr],
\end{eqnarray}
where $k_{j n} = \sqrt{k_{0}^2 \epsilon_{j} - n^{2} g^{2}}$ and $k_j\equiv k_{j0}$ (here $k_0=\omega/c$), and $r_{j n}$, and $t_{j n}$ are reflection and transmission amplitudes of order $n$ in medium $j$ respectively ($j = 1, 3$). In air/glass, the fields are expanded in the Bloch functions basis. In the medium 2 (perforated metal), the expansion goes over the normalized eigenstates of Maxwell equations (ME) in periodic medium, $H_i(x)e^{-\kappa_iz}$, where eigenvalue $\kappa_i$ is the wave-vector along the normal to the surface. The eigenfunctions $H_i(x)$ are localized in the hole region and are periodic over the array. The transverse electric fields in all three media are expressed through transverse magnetic fields via ME $E_{x} = -(i/k_{0} \epsilon) \partial H_{y} / \partial z$. In the relevant spectral range, it is sufficient to restrict to the lowest modes in air/glass ($n=0,\pm 1$), and in perforated metal $i=1$.

The transmission coefficient is obtained by matching transverse electric and magnetic fields at air/metal and metal/glass interfaces. It is convenient to perform this matching in the Bloch basis for electric field, and in the basis of eigenstates $H_i$ for magnetic field. This choice takes into account the fact that, for low volume fraction of holes, $\nu =\pi d^2/4a^2 \ll 1$, the electric field is mostly reflected back into the air at the air/metal interface and transmitted to the glass at the metal/glass interface, while the magnetic field, conversely, mostly penetrates into the metal  at the air/metal interface, and gets reflected back into the metal from  the metal/glass boundary. We define magnetic and electric field moments averaged over the Brilloin zone (BZ) as
\begin{eqnarray}
\label{bz}
&& h_{i n}=\langle e^{i n g x} H_i \rangle_{BZ},
\nonumber\\
&&e_{i n}=\langle e^{i n g x} E_i \rangle_{BZ} = \frac{i \kappa_i}{k_0} 
\biggl\langle e^{i n g x} \frac{H_{i}(x)}{\epsilon_{2}(x)}\biggr\rangle_{BZ},
\end{eqnarray}
where $\epsilon_2(x)=\epsilon_m+(\epsilon_h-\epsilon_m)\theta_H(x)$ is periodic dielectric function, $\epsilon_m$ ($\epsilon_h$) being dielectric function in metal (hole) region, and  $\theta_H$ is the periodic step-function vanishing outside the holes. Because we consider only the fundamental mode,  we drop index $i$ hereafter.
%
By performing matching procedure at the interfaces $z=0$ and $z=h$ we obtain the following system of equations 
\begin{eqnarray}
\label{bc} 
t_2  + r_2 u &=& h_{0}(1  + r_{1 0})  + 2 h_{1} r_{1 1}
\nonumber\\
e_{n} \,  (t_{2} - r_{2} u) &=&
\frac{k_{1 n}}{\epsilon_1 k_0} (\delta_{n 0}  - r_{1 n}) , 
\nonumber\\
t_{2} u + r_2&=& h_{0} t_{3 0} +2 h_{1} t_{3 1}, 
\nonumber\\
e_{n} \, (t_{2} u - r_{2}) &=&
\frac{k_{3 n}}{\epsilon_{3} k_0} t_{3 n},  
\end{eqnarray}
with $u=e^{-\kappa z}$, yielding the zero-order transmission
\begin{equation}
\label{t0}
T_0=|t_{3 0}|^2= \Biggl|\frac{t_{1 2} t_{2 3} e^{- \kappa h} }{1 + 
r_{1 2} r_{2 3} e^{- 2 \kappa  h} }\Biggr|^2,
\end{equation}
where $t_{1 2} = 2 \tilde{n}_{2} /( \tilde{n}_{1} + \tilde{n}_{2} )$, $t_{2 3} = 2 n_{3} /( \tilde{n}_{2} + \tilde{n}_{3} )$ are effective transmission amplitudes through the  air/metal, and metal/glass interfaces respectively, and  $r_{1 2} = (\tilde{n}_{2} - \tilde{n}_{1} )/( \tilde{n}_{1} + \tilde{n}_{2} )$, $r_{2 3} = (\tilde{n}_{3} - \tilde{n}_{2} )/( \tilde{n}_{2} + \tilde{n}_{3} )$ are the corresponding effective reflection amplitudes. The effective refraction indices $\tilde{n}$ are given by
\begin{eqnarray}
\label{n} 
\tilde{n}_{2} & =& \bigl(h_{0} e_{0}\bigr)^{-1}, 
\nonumber\\
\tilde{n}_{j}&  =& n_{j}\biggl(1 + \frac{2 h_{1} e_{1} k_{j}}{h_{0} e_{0} k_{j 1}}\biggr),
\end{eqnarray}
where $n_j$ are refraction indices in air/glass ($j=1,3$). The transmission coefficient is thus expressed via the {\em overlap integrals}, $h_n$ and $e_n$, between eigenstates in the periodic medium and in the air/glass. Such overlap causes hybridization of the states in perforated metal via the EM waves at the interfaces (both propagating and evanescent). It is this hybridization that determines the position and the lineshape of the transmission peak.

In order to proceed further, we need to evaluate the overlap integrals, $h_n$ and $e_n$, and the eigenvalue $\kappa$. The former can be determined from the following self-consistency argument. While in the bulk of perforated metal the states $H_i$ are strongly localized within holes, near the interfaces (where the overlap integrals are calculated) these states become delocalized due to the hybridization via EM in the dielectric. In other words, in the close vicinity of the interface, the profile of transverse magnetic field (that is continuous together with its derivative accross the interfaces) is similar at the both sides of the interface and can be approximated as $H(x) = h_0 + 2 h_1 \cos g x$. The relation between $h_0$ and $h_1$ on the metal side can therefore be determined from ME for $H(x)$,
\begin{eqnarray}
\label{me} 
\epsilon_2 \frac{\partial}{\partial x}\biggl(\frac{1}{\epsilon_2}\frac{\partial H}{\partial x}\biggl)
+\bigl(k_0^2\epsilon_2+\kappa^2\bigr)H=0,
\end{eqnarray}
with periodically modulated dielectric function, $\epsilon_{2}(x) = \epsilon_{0}  + 2 \epsilon_{1} \cos g x$, where $\epsilon_0 =  \epsilon_m (1 - \nu)+\epsilon_h \nu $ is the average and $\epsilon_1 =  \nu (\epsilon_h - \epsilon_m)\xi_1$, with  $\xi_n = \sin (ng d/2) / (ng d/2)$, is the modulation amplitude \cite{darmanyan03}.  In the first order in modulation, we obtain
\begin{eqnarray}
\label{dz} 
\frac{h_{1}}{h_{0}} = -\frac{2 k_{0}^2 \epsilon_{1}}{ k_{0}^2 \epsilon_{0} + \kappa^2 - g^2},
\end{eqnarray}
where $h_0$ is determined from the normalization condition. The moments of electric field, $e_n$, are straightforwardly calculated from Eq.\ (\ref{bz}) as
\begin{eqnarray}
\label{en} 
e_0 &=&  i \kappa (h_0 \tilde{\epsilon}_0 + 2 h_1
\tilde{\epsilon}_1)/k_0
\nonumber\\
e_1 &=& i \kappa \bigl[h_0 \tilde{\epsilon}_1 + 
h_1 (\tilde{\epsilon}_0 + \tilde{\epsilon}_2) \bigr]/k_0,
\end{eqnarray}
where $\tilde{\epsilon}_{n} \equiv \langle  e^{i  n g x}/ \epsilon_{2}(x) \rangle_{BZ} = (\epsilon_h^{-1} - \epsilon_m^{-1}) \nu \xi_n$.


Let us now turn to evaluation of  the wavevector $\kappa$ that represents the lowest eigenvalue of ME Eq.\ (\ref{me}). The eigenvalues are determined by the bulk states of the optically thick perforated metal film, and therefore are only weakly affected by the interfaces. In the bulk, the eigenstates are strongly localized within the holes and the weakly-modulated medium approximation does not apply. In this case, the eigenvalues can be evaluated by using the analogy between ME and Schr\"{o}dinger equation (SE) with complex potential. Namely, for each region, Eq.\ (\ref{me}) can be presented in the form $ \partial^2 H/\partial x^2+(E-V)H=0$, where 
\begin{eqnarray}
\label{se} 
&&
E=\kappa^2 +k_0^2 \bigl[\epsilon(\omega)-\epsilon_h \bigr],
\nonumber\\
&&
V(x) = k_0^2\sum_n\bigl[\epsilon(\omega)-\epsilon_h\bigr]\theta\bigl(d/2 -|x+na|\bigr),
\end{eqnarray}
are the energy and potential, respectively. The latter is Kronig-Penney type attractive potential of periodically spaced quantum wells. Note that  $E=-k_0^2\epsilon_h$ corresponds to $\kappa^2=-\epsilon_mk_0^2$ that is the value of $\kappa$ for a solid metal film. The fundametal mode is associated to the ground state of Eq.\ (\ref{se}) with the energy spectrum being a band with the width determined by the tunnel transparency between the wells. In the case of high metal/dielectric contrast of $\epsilon$, so that $k_0a\sqrt{|\epsilon_h-\epsilon_m|}\gg 1$, this transparency is exponentially small and the ground state energy can be approximated by that of the lowest level in a single well. The latter can be found by replacing rectangular well by the delta-function potential, $V(x)=V_0\delta(x)$, with the strength $V_0=\nu (\epsilon_m-\epsilon_h) k_0^2a$, where we averaged over BZ.  We then obtain $E=-V_0^2/4$, yielding $\kappa^2= -\epsilon_mk_0^2 - \nu^2  a^2 k_0^4 (\epsilon_m-\epsilon_h) ^2/4$.

We thus arrive at the following physical picture for the main transmission band. Diffraction on an individual hole can be viewed as scattering on the bound state in the associated SE. The interference of diffracted waves, caused by the hybridization of localized hole states via EM waves in dielectrics, leads to the single-period modulation of magnetic field profile at the interfaces. The latter is determined by the {\em effective medium} characterized by weakly-modulated  dielectric function depending on the hole volume fraction $\nu$. Note that eigenvalue $\kappa$ as well as  magnetic/electric field moments $h_n, e_n$ are complex due to absorptive part of metal dielectric function $\epsilon_m=\epsilon'_m+i\epsilon''_m$.

In Fig.\ \ref{fig:1} we show the fit of linear trasmission obtained using the Drude form of dielectric function of Al, $\epsilon_m$, with bulk plasmon frequency $\omega_p = 15.0$ eV and damping rate $\gamma = 0.2$ eV \cite{palik}. It can be see that the main transmission band is described accurately, including the position of Wood's anomaly at $\omega \approx 2.6$ eV. Higher order maxima, including possible SPP peak at $4.0 $ eV are not captured within this model.

\subsection{Ultrafast Dynamics}
\label{sec:ultrafast}

We now turn to the modeling of ultrafast dynamics. The transient spectra in a perforated film can be obtained from  the linear transmission $T_0(\omega)$ by the replacement $\varepsilon_{m}(\omega) \rightarrow \varepsilon_{m}(\omega) + \delta\varepsilon_{m}(\omega, t)$, where $\delta\varepsilon_{m}(\omega, t)$ is the non-equilibrium contribution to the metal dielectric function. As noted above,  $\delta\varepsilon_{m}(\omega, t)$ is mainly due to transient change in the {\em e-ph} relaxation rate $\gamma(t)$ in the Drude dielectric function. The distinctive feature of electron dynamics in Al films is a very fast (hundreds of fs) energy exchange between electron and phonon subsystems caused by strong {\em e-ph} interaction in Al. This energy exchange  takes place on the same timescale as the electron gas thermalization due to {\em e-e} scattering, rendering the transient dynamics inherently non-equilibrium.  Note however that the high-energy fraction of of phoexcited electrons with the excess energy $(E-E_F) \sim \hbar\omega$, scatter down on the timescale of tens of fs due to the quadratic energy dependence of {\em e-e} scattering rate, $\gamma_{ee}\propto (E-E_F)^2$, whereas the thermalization of electron population is reached on the longer timescale of hundreds of fs \cite{sun94}. Therefore, the higher energy end of electron distribution  relaxes primarily via {\em e-e} scattering processes while the energy transfer to the phonons primarily goes through the lower energy electrons. Since the overall heating of the lattice is of the order of Debye temperature ($\Theta_D \approx 380 K$ in Al), the phonons that gain energy from athermal electron gas can be considered as quasi-equilibrium. The energy exchange between phonon and electron subsystems can be then described in terms of time-dependent latice temperature $T_L(t)$ as  
\begin{equation}
C_L \frac{\partial T_L}{\partial t}  
= 
- \int dE g(E) E \biggl(\frac{\partial f}{ \partial t} \biggr)_{e-ph} ,
\label{lattice} 
\end{equation}
where $C_L$ is the lattice heat capacity ($C_L= 900 $ J/Kkg for Al), $g(E) = E^{1/2} (2 m_e)^{3/2}/2 \pi^2 \hbar^3$ is the electron density of states (DOS), and  $f(E,t)$ is the electron distribution function that satisfies the Boltzmann equation
\begin{equation}
\label{boltzmann}
\frac{\partial f}{ \partial t} = 
\biggl(\frac{\partial f}{ \partial t} \biggr)_{e-e} 
+ \biggl(\frac{\partial f}{ \partial t}\biggr)_{e-ph} 
+ H(E,t).
\end{equation}
Here the first two terms in r.h.s. are due to {\em e-e} and {\em e-ph} scattering processes respectively, and $H(E,t)$ stands for electron-gas excitation by the pump pulse. Naturally, only the  {\em  e-ph} scattering term contributes to the energy exchange rate Eq.\ (\ref{lattice}). Since the typical phonon energy is much smaller than the electron energy, the {\em  e-ph} term can be more conveniently presented in the differential operator form \cite{fatti00}. In the deformation potential approximation, the  {\em  e-ph} scattering term reads 
\begin{equation}
\left(\frac{\partial f}{ \partial t}\right)_{e-ph}  = 
\frac{C_{ep}}{\sqrt{E}} \frac{\partial}{\partial E} 
\left[ 
f \left(1-f\right) + k_{B} T_{L}(t) \frac{\partial f}{\partial E} 
\right],
\label{diffe-ph}
\end{equation}
where $C_{ep} = \pi^2 \hbar^3 G/\sqrt{2}k_B m_e^{3/2} $ and $G$ is {\em e-ph} interaction constant. By substituting Eq.\ (\ref{diffe-ph})  into (\ref{lattice}) we obtain the rate equation for the lattice temperature,
\begin{equation}
C_L \frac{\partial T_L}{\partial t} = G (T^{*}_e - T_L) +S_{GL}(t),
\label{2tphonon}
\end{equation}
where  
\begin{equation}
T_{e}^{*}(t) = \int dE f(E,t) \bigl[1 - f(E,t)\bigr]
\label{Teff}
\end{equation}
plays the role of {\em effective} temperature of {\em athermal} electrons. Note that for thermal electron distribution $f_0$, $T_{e}^{*}$ coinsides with the ``regular'' electron temperature $T_{e}$. The last term in Eq.\ (\ref{2tphonon}) is added to describe heat exchange between the metal film and the glass substrate. With boundary conditions that the temperature and heat flux are continuous accross the metal/glass interface, it has the form \cite{vallee03}: $S_{GL}(t)=\kappa_G\, \partial T_G/\partial z$, where $T_G$ and $\kappa_G$ are the glass temperature and thermal conductivity, respectively ($\kappa_G=1.37$ J/mK  for SiO$_2$). The above system should be supplemented by equation for heat transport in the glass, 
\begin{eqnarray}
C_G {\partial T_G \over \partial t} = \kappa_G  \frac{\partial^2 T_G }{\partial z^2},
\label{heat}
\end{eqnarray}
where $C_G $ is the glass heat capacity ($C_G =1.63 \times 10^7$ J/m$^2$K for SiO$_2$). Note, however, that metal/glass dynamics is important only for long times (see below).

In the case short-time dynamics, the evolution of lattice temperature $T_L(t)$ is governed by the interaction with strongly athermal electrons gas, and thus requires solution of Boltzmann equation for {\em e-ph} system (\ref{boltzmann}-\ref{2tphonon}). We use the integral  form for the {\em e-e} scattering term for parabolic band \cite{fatti00},
\begin{eqnarray}
\biggl(\frac{\partial f}{ \partial t}\biggr)_{e-e} 
&=& 
\frac{K_{ee}}{E_s^{3/2} E^{1/2}} \int dE_1 \int dE_2 \,
F(E_{1}, E_{2}, E_{3})  \nonumber \\
&&\times 
\left[ \frac{\sqrt{x}}{1+x} + \arctan \sqrt{x} 
\right]_{x_{min}}^{x_{max}},
\label{dfdtee}
\end{eqnarray}
where $K_{ee} = m_e e^4 / (4 h^3 \varepsilon^2)$, $\varepsilon$ is static dielectric constant of metal, and $E_s = (e^2 /2 \pi \varepsilon )\bigl( 3 n_e/\pi \bigr)^{1/3}$ is the Thomas-Fermi energy. The phase space for scattering is determined by $F(E_1, E_2, E_3) = \tilde{f}(E) \tilde{f}(E_1) f(E_2) f(E_3) - f(E) f(E_1) \tilde{f}(E_2) \tilde{f}(E_3)$, where $\tilde{f}(E) = 1 - f(E)$ and $E_3 = E + E_1 - E_2$, while the angular integrals of Coulomb interaction yield the expression in the brackets in the integrand which is evaluated between the  limints: $x_{min, max} = \mbox{max,min} \Bigl[\bigl( \sqrt{E_1} \mp \sqrt{E_3}\bigr)^{2}, \bigl( \sqrt{E} \mp \sqrt{E_2}\bigr)^2\Bigr] /E_s$. The last term in Eq.\ (\ref{boltzmann}), describing the energy deposited by the pump pulse, is given by $H(E,t)= Q(E)S(t)$ with Gaussian pulse profile
\begin{eqnarray}
S(t) =  \frac{E_0}{\sqrt{\pi} \sigma_L} \, e^{-t^2/\sigma_L^2},
\end{eqnarray}
where $E_0$ is the absorbed pump pulse energy density and $\sigma_L$ = 100 fs is the pulsewidth. The factor $Q(E)$, describing the phase space of phonon-assisted electronic transitions after absorption of a photon with energy $\omega$, is available in the literature \cite{fatti00} and is included in the numerical calculations.

\begin{figure}
\centering
\includegraphics[width=2.5in]{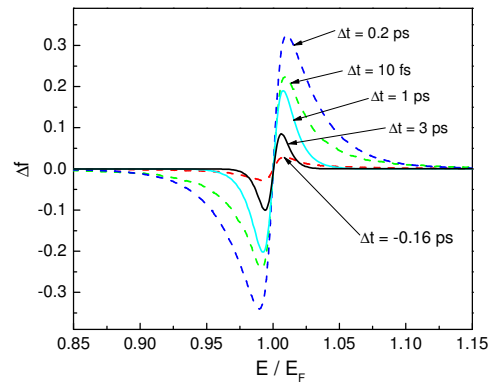}
\caption{ 
Non-equilibrium contribution to the electron distribution function in Al film is shown for several time delays. Calculations are based on solution of the Boltzman equation for the electron distribution function in Sec. \ref{sec:ultrafast}.
} 
\label{fig:5}
\end{figure}

In Fig.\ \ref{fig:5}, we plot the time-dependent contribution to the electron distribution function, $\Delta f (E,t)=f(E,t)-f_0(E)$,  versus energy (measured from the Fermi level) at several time delays. For short times, electron distribution is strongly non-equilibrium, as indicated by the long high-energy tail of $\Delta f$. At $t\sim 1$ ps, electrons are almost thermalized and their distribution is close to the Fermi function of a hot electron gas. Note that, at 1 ps, the magnitude of $\Delta f$ has decreased significantly pointing to the energy transfer from electron gas to the lattice.
\section{Discussion}

The transient change in the lattice temperature, $T_L(t)$, provides the main contribution to the time-modulation of metal dielectric function,
\begin{eqnarray}
\label{pm}
\delta\varepsilon_m (\omega,t)=
\frac{i\omega_p^2}{\omega (\omega+i\gamma_0)^2}\,
\bigl[\gamma \bigl(T_L(t)\bigr)-\gamma_0\bigr],
\end{eqnarray}
where $\gamma (\omega,T_L)$ is the optical relaxation rate of an electron-hole pair excited by the probe light \cite{gotze&wolfle72} ($\gamma_0$ stands for the room-temperature relaxation rate). In the case when the lattice temperature remains much smaller than $\omega$, the relaxation rate is nearly $\omega$-independent \cite{smith82}, and is given by 
 %
\begin{equation}
\label{J(y)}
\gamma(T) =
 B \biggl(\frac{T_L}{\Theta_D}\biggr)^5 J\biggl(\frac{\Theta_D}{T_L}\biggr),
 \end{equation}
where $J(y) = \frac{1}{2} \int_0^y dx x^4 \mbox{coth}\frac{x}{2}$, and $B$ is a material-dependent constant. In the relevant  temperature range $ T_L \sim \Theta_D$, the $T_L$--dependence of $\gamma$ is nearly linear, $\gamma(T_L) \approx \gamma_0 + \frac{B}{2}(T_L-T_0)$. Note that $\gamma_0$ contains contributions also from the electron-impurity and {\em e-e} scattering processes; however those are not significantly affected by the pump pulse due to a relatively small change in the scattering phase space by the fraction of photoexcited electrons.    

\begin{figure}
\centering
\includegraphics[width=2.5in]{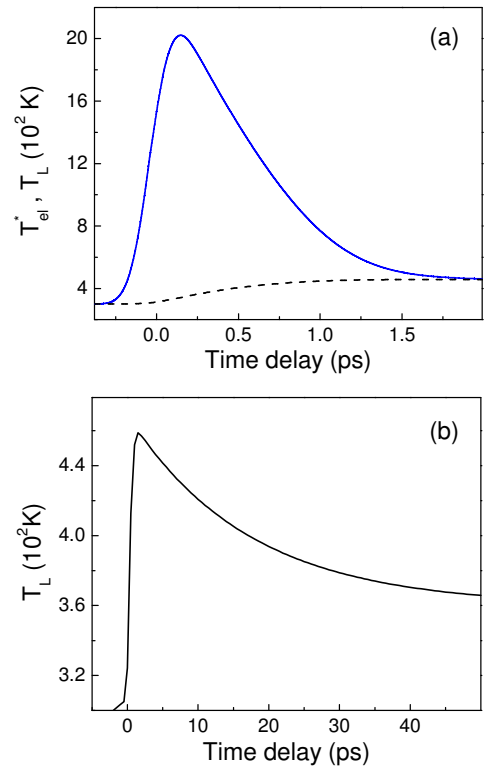}
\caption{(a) Short-time evolution of effective electron (solid line) and lattice (dashed line) temperatures in Al film following the excitation. (b) Long-time evolution of the lattice temperature due to heat transfer to the glass matrix.
} 
\label{fig:6}
\end{figure}

The results of numerical simulations of short-time dynamics are shown in Fig.\ \ref{fig:2}. The calculated differential transmission $dT/T$ is in excellent agreement with the data, reproducing the signal rise at subpicosecond times. In Fig.\ \ref{fig:6}  we show the evolution of the lattice temperature and the effective electron temperature. The rapid drop of $T_e^*$ after reaching its peak value at $\sim 200$ fs is accompanied by increase in $T_L$ during the period $t \lesssim 1$ ps, indicating energy intake by the phonon subsystem while the electron gas is still in the thermalization stage. After electron and phonon subsystems equilibrate, the signal changes weakly (plateau) until the slower process of energy transfer to the glass substrate takes over, leading to the decay of pump-probe signal from the metal [see Fig.\ \ref{fig:3}]. For long time delays, electron and phonon subsystems are in thermal equilibrium with each other, and the long-time dynamics is well descrbed by the \textit{two-temperature model} that incorporates the heat exchange between the metal and the glass \cite{vallee03}. To compare to Boltzmann equation approach, here we also included in calculations the rate equation for electron temperature $T_e(t)$ that can be straightforwardly obtained from Eqs.\ (\ref{boltzmann}-\ref{2tphonon}) by the replacement of $f(E,t)$ with thermal distribution: 
\begin{equation}
\label{2T}
C_{e}\, \partial T_{e}/ \partial t = G (T_L - T_e) + S(t), 
\end{equation}
where $C_e = \beta T_e$ (with $\beta = 134$ J m$^{-3}$ K$^{-2}$ for Al) is the electronic heat capacity. After $\sim 40$ ps, the signal reaches a plateau corresponding to thermal equilibrium between film and glass [see Fig.~\ref{fig:3}]. Here the evolution of $dT/T$ also follows mainly the phonon dynamics [see Fig.~\ref{fig:6}(b)]. Note, however, that the subpicosecond signal rise is noticably faster than the one calculated using Boltzmann equation approach [cf. Fig.\ \ref{fig:2}]. The reason is that thermal distribution overestimates the fraction of  lower energy electrons and, hence, the rate of their energy transfer to the lattice. The measured signal also exhibits weak oscillations with a period $\sim 5$ ps; those could be attributed to the vibrational mode accross the film that is excited by the rapid heating of the lattice, but are not captured by the present model. 

The calculated transient spectra are shown in  Fig.\ \ref{fig:4} for several time delays in the long-time regime when the electron and phonon subsystems are in thermal equilibrium with each other. The blueshift is well reproduced as well as is the nonlinear $\omega$-dependence of $dT/T$ in the main transmission band frequency range. The origin of the blueshift can be understood as follows. The eigenmodes in a nanohole array are the result of interference of multiple waves scattered from periodic metal/dielectric regions as described by the above effective medium model. In the presence of \textit{attenuation}, the  phase of scattered wave has a contribution $\propto \mbox{Re} \sqrt{\varepsilon_{m}} = 	\varepsilon_{m}''/\sqrt{|\varepsilon_{m}'|} \approx \gamma \omega_p/\omega^2$. Therefore, a transient change in attenuation causes corresponding shifts in scattering phases. This, in turn, changes the interference of waves making up an eigenmode. To compensate for the phase shifts, a transient increase in $\gamma$ must be accompanied by a corresponding increase in $\omega$; and consequently the transmission resonances blueshift.

\section{Conclusion}
In summary, we investigated the ultrafast dynamics of Al-based plasmonic lattice that reveals a non-trivial role played by the plasma attenuation in determining the EOT resonant bands. While its effect on the SPP dispersion is relatively weak, attenuation is more important for modes comprised of a multitude of scattered waves. By directly altering their interference, ultrafast time-resolved optical techniques unravel the unique properties of photoexcited SEW in plasmonic metals. 

This work was supported in part by NSF DMR Grant No. 05-03172,  by NSF Grant No. DMR-0606509, and by NIH Grant No. 2 S06 GM008047-33.

\end{document}